\documentstyle[twoside,fleqn,espcrc2,epsf]{article}

\newcommand{\AmS}{{\protect\the\textfont2
  A\kern-.1667em\lower.5ex\hbox{M}\kern-.125emS}}

\title{
  \vspace{-1.5cm}
\hfill \normalsize{OCHA-PP-141} \\
\vspace*{0.3cm}
    Genetic Algorithm for Lattice Gauge Theory. \\
   \normalsize{On SU(2) and U(1) on 4 dimensional lattice,
   how to hitchhike to thermal equilibrium state .}
}
\author{
 A. Yamaguchi 
 and A. Sugamoto  \thanks{ talked by A.Sugamoto}  \\
{Particle Physics Lab.
           Department of Physics, 
           Ochanomizu University, Tokyo, Japan }}%
\begin{document}

\begin{abstract}
Applying Genetic Algorithm for the Lattice Gauge Theory is formed to be an
effective method to minimize the action of gauge field on a lattice.  In 4
dimensions, the critical point and the Wilson loop behaviour of SU(2) lattice
gauge theory as well as the phase transition of U(1) theory
have been studied. The proper coding methodi has been developed in
order to avoid the increase of necessary
memory and the overload of calculation for Genetic Algorithm.
How hichhikers toward equlibrium appear against kidnappers is clarified.

\end{abstract}

\maketitle

\begin{figure*}[tb]
\begin{center}
\begin{tabular}{cc}
\epsfxsize=7.5cm\epsffile{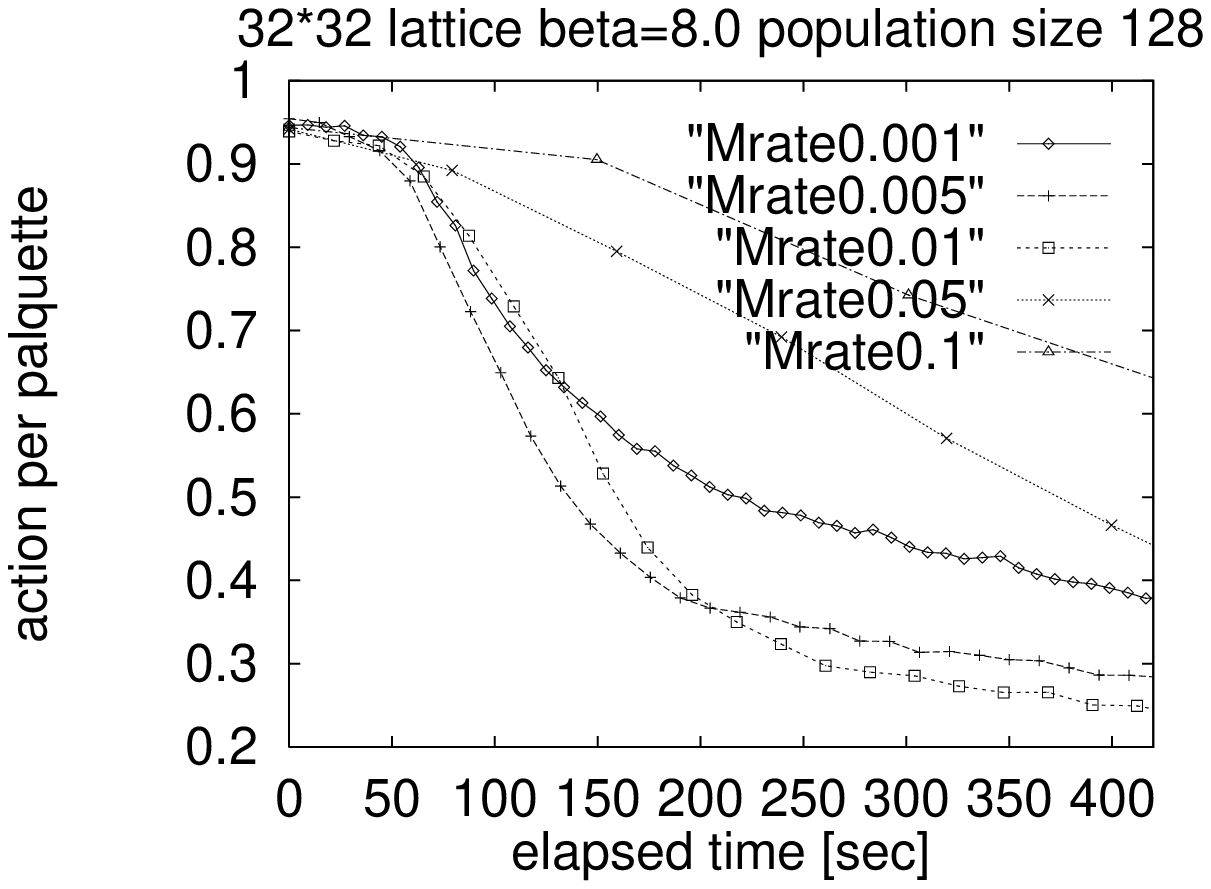} &
\epsfxsize=7.5cm\epsffile{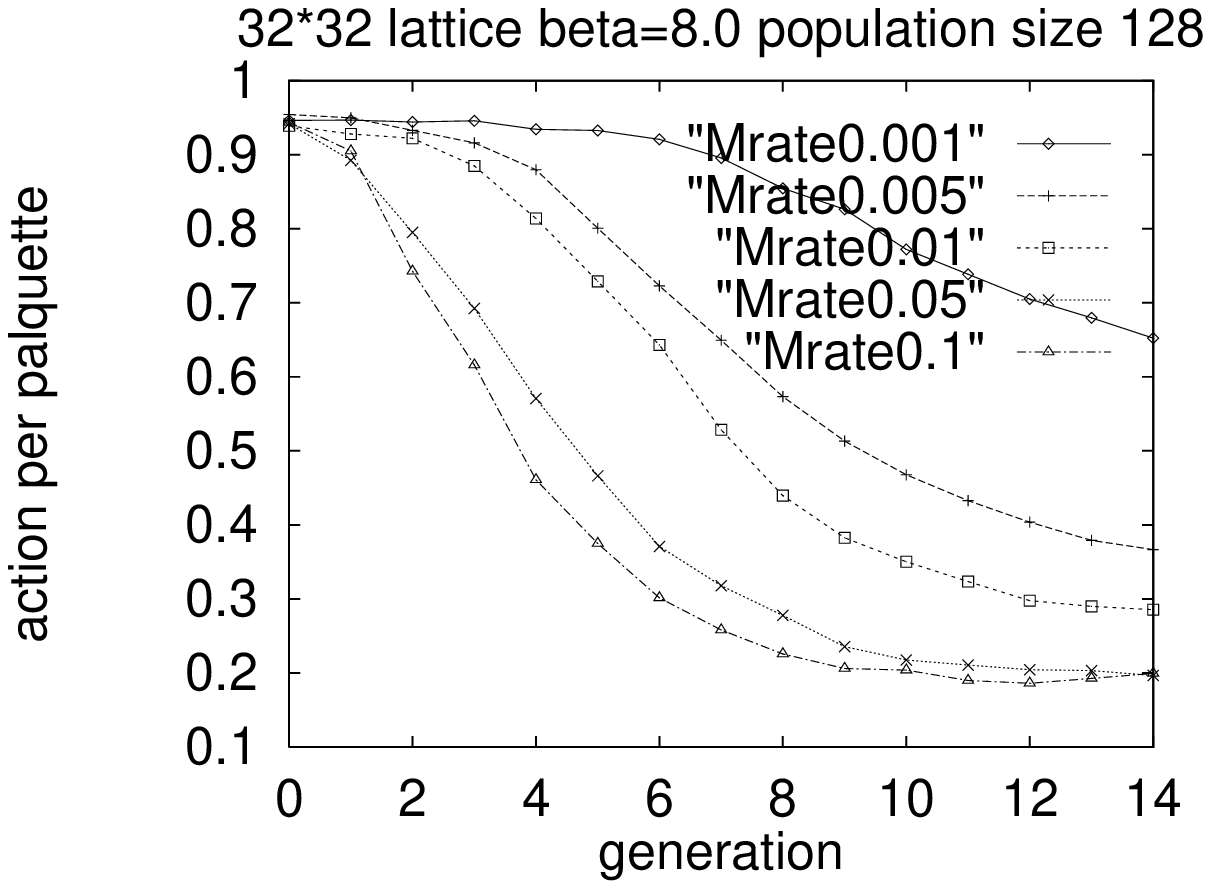} \\
\hspace{0.3cm}
The horizontal axis is computing time.&
\hspace{0.3cm}
The horizontal axis is the number of generation.\\
\end{tabular}
\end{center}
\vspace{-1.0cm}
\caption{ \label {fig1}
 Thermalizaion of Action per Plaquette
  $32 \cdot 32$ lattice $\beta=8.0$ , population size 128}
\end{figure*}%

The goal of numerical simulation is 
to get the maximum (minimum) value of a given 
objective function parametrized by some set of independent variables.
A genetic algorithm originated by J.H.Holland ~\cite{Hol} 
is known that it is good at searching
the maximum/minimum value under complicated situations.
It bases its search on a set of point/individual of a searching space and
on selection with respect to its fitness value, that is,
the search of genetic algorithm is a global one rather than those of Metropolis
and Heatbath method.
Some severe extra works for coding a problem toward an
appropriate form and treating some number of individuals at once had
made its application wait until operating ability of computer system
becomes adequate. By the recent development of the computer systems,
its wide implementation becomes possible.
However, it is still a subtle point that how to code a genome and how to 
design operations for the sake of good performance .
We develop the coding of genomes in the SU(N) and U(N) lattice
gauge theories and investigate to tune its 
searching parameters.

\section{Algorithm and Coding}

A configuration on a lattice is treated as a phenotype.
Link variables on a site are taken from the link variable pool 
created in advance by the integer number index 
which is coded to a gene, a binary string.
In this coding, the memory for genetic algorithm operation 
is able to be suppressed as follows;
With the number of link variables in a link pool, $2^{Lp} $,
the lattice size,  $2^{Lx+Ly+Lz+Lt}$, and the population size,
$2^{pop\_size} $, the length of a genome becomes
$ Lp \times n\_dim \times $ lattice size.

For instance, for $8\time8\time8\time8$ SU(2) lattice,
 a genome( a lattice) needs $282 kb$, and  a population (32 lattices)
needs about $9.2Mb$.
On the other hand, but for the genetic algorithm, the same size lattice
 needs 1.5 Mb and 32 lattices would need  $48 Mb$. 
That is, a genetic operation needs only a quarter of memory for
other methods.

Our algorithm has three genetic operations;
besides ordinal ones of selection and recombination, 
a new one we call "the intergenerational conflict" is
introduced to establish the thermal equilibrium. 
Furthermore as a local search, Metropolis updating applied to a link variable 
on a lattice is added, with which 
our combined algorithm behaves like a hybrid genetic algorithm.

This extraneous operation, the intergenerational conflict,
is applied to the parent with a smaller action between a pair of parents'
genomes and the offspring with a smaller action
 between the pair of offsprings produced from the parents.
Offspring is passed to the next generation  under the condition expressed
using an ordinary Metropolis function,
\begin{eqnarray*}
&min\{1,exp^{-\beta(S_{o} -S_{p})} \} > \xi, & \\
\mbox{where}& & \\
& \displaystyle{
  S = S[U] = \beta\Sigma_p \left( 1 - \frac{1}{N}ReTrU_p \right),
             } & \\
& \xi \ \ \mbox{ an uniform random number.} &
\end{eqnarray*}
Otherwise the parent is passed.
Metropolis updating is applied to the survivor of 
the intergenerational conflict.

All genetic
operations are coded in bit calculations. Arithmetic operation which needs
more executed time is used for the calculation of the action. 
This coding  can reduce the increasing of the executing time. For instance, 
genetic operations for $8^4$ lattice with $\beta=2$ occupies 17 percent 
of the executing time for one cycle, 
that is the increasing time caused by genetic operations is
15 percent of Metropolis method.
This analysis conditions 
our genetic algorithm to an advantage against the usual
Metropolis method.

\begin{figure}[tb]
\begin{center}
\epsfxsize=7.5cm\epsffile{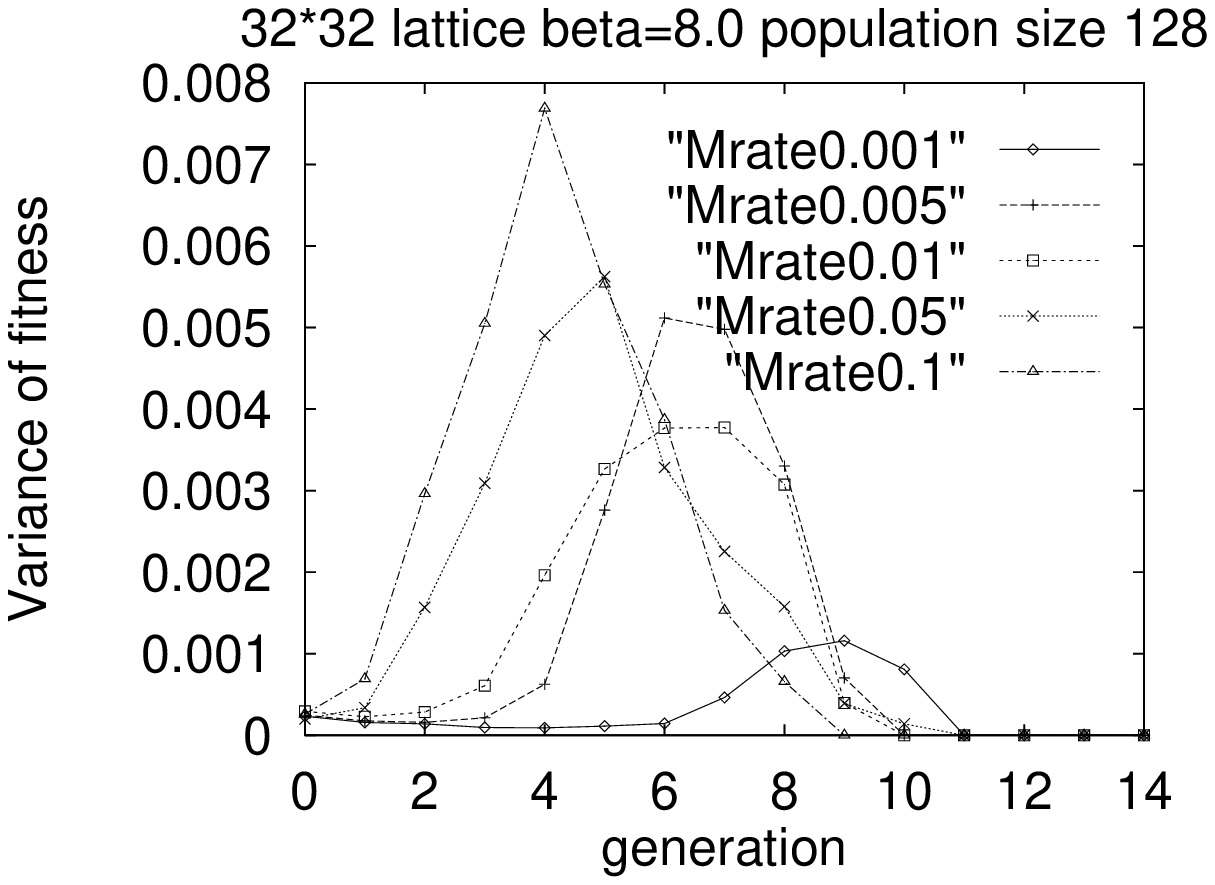}
\end{center}
\vspace{-1.0cm}
\caption{ \label {fig2}
 The variance of fitness,
  $32 \cdot 32$ lattice $\beta=8.0$ , population size 128}
\vspace{-1cm}
\end{figure}%

\section{The algorithm dynamics}

To discuss the profitable condition, 
the dynamics of our algorithm on the statistical point of view
is analized.
For the procedure of Metropolis updating on a lattice is 
a random selection of a link $\Rightarrow$ decoding information of gene 
$\Rightarrow$ fixing the index of the link variable 
$\Rightarrow$ updating the variable.  
This Metropolis updating makes configurations converge fast or
slow and or not converge at all, because it work not only
on the target lattice but also on the other lattice in the stack.
This updating duplication makes two cases occur. 
First one is that if Metropolis updating works well even
for another lattice configuration, the other lattice's  action decreases. 
That is, its fitness value increases, and the lattice which is not
the target of the Metropolis updating does hitchhike. 
Second one is the opposite case, that is,
if Metropolis updating works well for the target lattice configuration
but for the other lattice in the stack, its action increases that is
its fitness value decreases. In this case, the target lattice kidnaps the
other one. For the kidnapped lattice, the probability to be selected as 
 a parent becomes low.

\begin{table}
\caption{ 
 U(1) Action on 4 dimension \label{tab:U1} }
\begin{tabular}{lll}
$\beta$ & Metropolis & GA \\
$1.0$ & 0.538  & 0.538 \\
$1.5$ & 0.370 & 0.369 \\
$2.0$ & 0.298 & 0.292 \\
$3.0$ & 0.237 & 0.235 
\end{tabular}

\caption{ 
SU(2) Action on 4 dimension \label{tab:SU2} }
\begin{tabular}{llll}
$\beta$ &  Metropolis & GA \\
$2.0$ &  0.565 & 0.564 \\
$4.0$ &  0.384  & 0.34   
\vspace{-.5cm}
\end{tabular}
\end{table}
\begin{figure}[tb]
\vspace{-1.0cm}
\begin{center}
\epsfxsize=7.0cm\epsffile{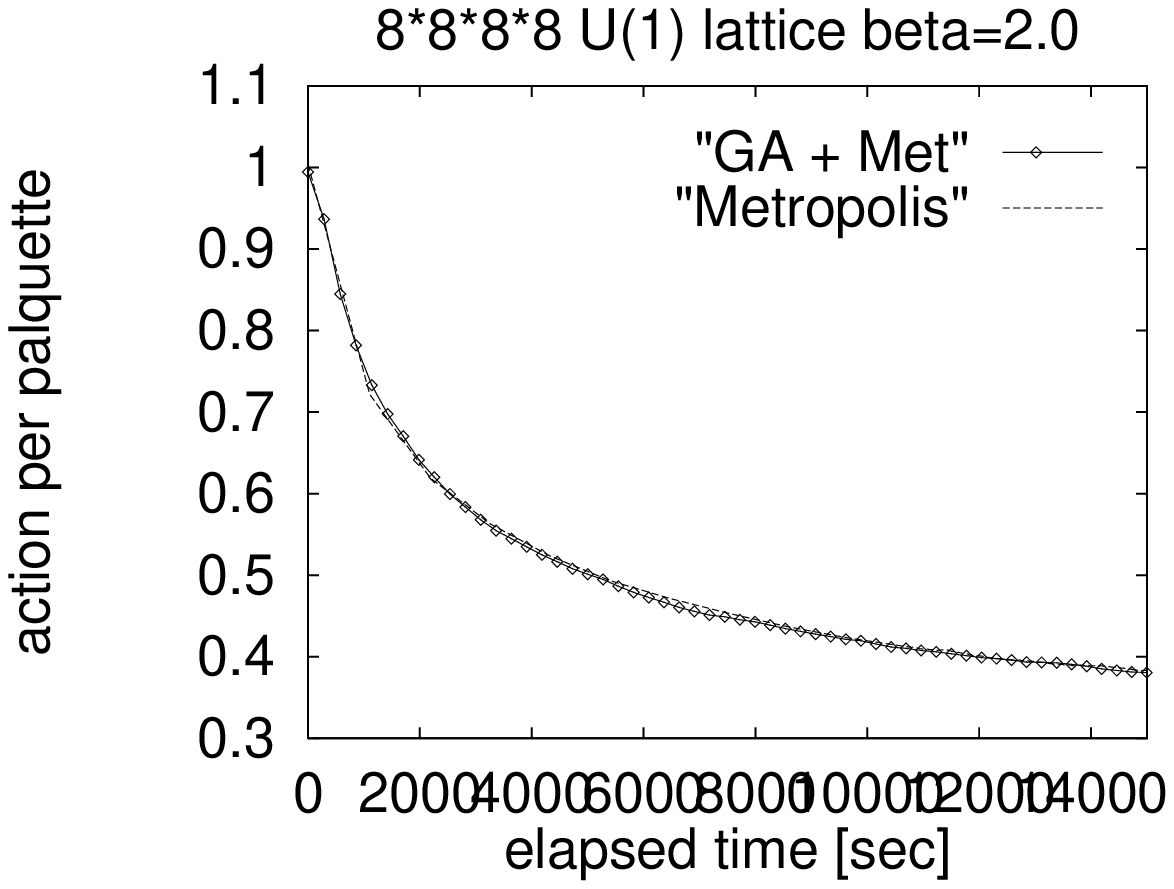}
\end{center}
\vspace{-1.5cm}
\caption{ \label{fig:U1A}
Thermalization of the U(1) action on 
$8^4$ lattice $\beta=2.0$, GA population size 16
  and Metropolis method}
\vspace{-1.0cm}
\end{figure}%

The behaviour of  Hitchhiker and Kidnaper are 
clarified from the statistical value of the fitness.
Five different Metropolis updating ratios, from 0.1 percents to 10 percents are
set. From Fig.~\ref{fig1}, the Metropolis updating ratio is found to depend 
strongly on the performance, so that the Metropolis updating ratio should 
be set 1 percent and not 5 percent nor 0.5 percent.

In Fig.~\ref{fig1} with generation, 
the effect of kidnapers and hitchhikers are appeared.
There is not so clear difference between 10 percent line and 5 percent line.
The slowing down of 10 percent line is caused by the kidnapers. 
On the other hand, the line of 1 percent and 5 percent fall rather fast, 
which shows there are some hitchhikers in the populations. 
From the variance of fitness, 
the populations which have some hitchhikers keep the variance
at some level not so high and not so low. It means that some lattices might
be kidnapped but other hitchhiking lattices accelerate its thermalization speed.
It means the ratio of Metropolis updating has to be tuned.
We set $0.01$ for $\beta = 8.0$ and $0.005$ for $\beta = 2.0$

\section{ Result and Discussion }
The physical values obtained by our algorithm are consistent with those given
by Metropolis. See Tab.~\ref{tab:U1} and ~\ref{tab:SU2}.

The mechanism Hitchhiker and Kidnaper is
clarified from the statistical value of the fitness.
Since the Metropolis updating ratio depends 
strongly on the performance, it should 
be set 1 percent not 5 percent nor 0.5 percent to $\beta=2.0$.

Without any tuning parameters, a performance of our algorithm
for U(1) on 4 dimensional lattice with 16 population size and $\beta=2.0$,
is same as that of Metropolis method. See Fig.~\ref{fig:U1A}.

For SU(2), on 2 dimensional lattice, 
our algorithm can get higher 
performance than Metropolis method~\cite{AZ}. On 4 dimensional lattice, however,
it's performance is about the  same as that of Metropolis.
We have not tuned the parameter for hitchhikers on 4 dimensional
lattice yet, but the performance is strongly dependent on 
the Metropolis updating ratio, because it occupies calculation time
mainly.

Our work suggests that an algorithm with genetic operation does 
work in the case that the thermal equilibrium state is to be
established. 

The introduction of fermion must follow as the next stage,
since our coding is good for SU(3) lattice gauge theory.

\section{Acknowledgment}
We acknowledge the use of the Reproduction Plan Language, RPL2 produced
by Quadrastone Limited, and
the use of workstations of Yukawa Institute and Kitano Symbiotic Systems 
Project.

\end{document}